\documentclass[12pt]{article}
\usepackage{pst-plot,url,epsf}
\newcommand{\beq}{\begin{equation}}
\newcommand{\eeq}{\end{equation}}
\newcommand{\bea}{\begin{eqnarray}}
\newcommand{\eea}{\end{eqnarray}}
\newcommand{\beas}{\begin{eqnarray*}}
\newcommand{\eeas}{\end{eqnarray*}}

\newcommand{\epm}{e^+e^-}

\newcommand{\ra}{\rightarrow}

\newcommand{\AmS}{{\protect\the\textfont2
  A\kern-.1667em\lower.5ex\hbox{M}\kern-.125emS}}

\newcommand{\nn}{\nonumber}

\begin{document}
\thispagestyle{empty}
\hspace*{14.cm} March 2009\\
\hspace*{14.cm} Revised:\\
\hspace*{14.cm} September 2009\\[1.5cm]
\begin{center}
{\LARGE\bf $\epm \ra t \bar t H$ including decays:\\[2mm] on the size
of background contributions}\\
\vspace*{2cm}
Karol Ko\l odziej\footnote{Corresponding author.\\
{\em E-mail addresses:} karol.kolodziej@us.edu.pl (K. Ko\l odziej),
simon\_t@poczta.fm (S. Szczypi\'nski)} 
and Szymon Szczypi\'nski\\[6mm]
{\small\it Institute of Physics, University of Silesia\\ 
ul. Uniwersytecka 4, PL-40007 Katowice, Poland}\\
\vspace*{2.5cm}
{\bf Abstract}\\
\end{center}
We present results for the lowest order cross sections, calculated
with the complete set of the standard model Feynman diagrams, of all possible
detection channels of the associated production of the top quark pair and 
the light Higgs boson, which may be used for determination 
of the top--Higgs Yukawa coupling at the future $\epm$ linear collider. 
We show that, for typical particle identification cuts, the 
background contributions are large.
In particular, the QCD background contributions are much bigger than 
could be expected when taking into account a possibly
low virtuality of exchanged gluons. Moreover, we include the initial state
radiation effects and discuss the dependence of the cross sections
on the Higgs boson and top quark masses.
\vfill

\newpage

\section{Introduction}

A reaction of associated production of a top quark pair and a Higgs boson
at the future $\epm$ linear collider 
\begin{equation}
\label{eetth}
 \epm \ra t \bar{t} H
\end{equation}
can be used to determine the top--Higgs Yukawa coupling \cite{eetth}.
In the Standard Model (SM), after taking into account decays of the top quark,
$t\ra bW^+$, of the antitop quark, $\bar t\ra \bar bW^-$, together with 
the subsequent decays of the $W$ bosons, and of the Higgs boson that,
if its mass is less than about $140$~GeV, decays preferably into
a $b\bar b$-quark pair, reaction (\ref{eetth}) takes the following form
\beq
\label{ee8f}
  e^+e^-\;\; \ra \;\;  b \bar b  b \bar b f_1\bar{f'_1} f_2 \bar{f'_2}.
\eeq
If we neglect the Cabibbo--Kobayashi--Maskawa (CKM) mixing in (\ref{ee8f})
then $f_1, f'_2 =\nu_{e}, \nu_{\mu}, 
\nu_{\tau}$, $u, c$ and $f'_1, f_2 = e^-, \mu^-, \tau^-, d, s$, respectively.
For the heavier Higgs boson, with mass $m_H > 140$~GeV, the decay
into electroweak boson pair will dominate and the detection channels
of (\ref{eetth}), will heave 10 fermions in the final state. 
This option is much more complicated and it will not be addressed in 
the present work.

\begin{figure*}[htb]
\vspace{4cm}
\includegraphics{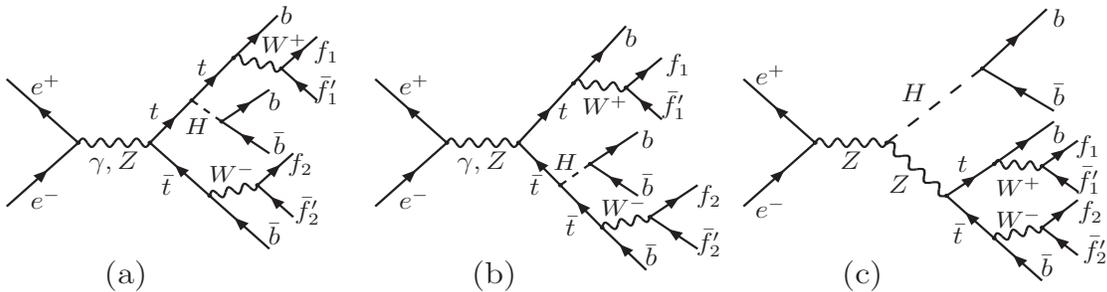}
\caption{Representative signal Feynman diagrams of reaction (\ref{ee8f}) in 
the unitary gauge.
The remaining diagrams are obtained by 4 permutations of 
the two $b$ and two $\bar b$ lines. The Higgs boson coupling to electrons
has been neglected.}
\label{fig:ee8f}
\end{figure*}

Reactions (\ref{ee8f}) receive contributions from tens thousands of Feynman 
diagrams already in the lowest order of SM. 
For example, reactions
\bea
\label{tnmn}
\epm &\ra& b \bar b b \bar b \nu_e e^+ \mu^- \bar \nu_{\mu},\\
\label{udmn}
\epm &\ra& b \bar b b \bar b u \bar d \mu^- \bar \nu_{\mu},\\
\label{cssc}
\epm &\ra& b \bar b b \bar b c \bar s s\bar c,
\eea
which correspond to the leptonic, semileptonic and hadronic decays of
the $W$ bosons, receive contributions, respectively, from
37\,868, 26\,816 and 240\,966 Feynman diagrams in the unitary gauge,
with the neglect of the Yukawa couplings
of the fermions lighter than the $c$ quark.
An overwhelming majority of 
the diagrams comprise background to the resonant production 
and decay of the top quark pair and Higgs boson that is mediated by
20 signal Feynman diagrams which contain 
the propagators of the top, antitop and Higgs at a time.
The representative signal Feynman diagrams  are shown in
Fig.~\ref{fig:ee8f}.

To which extent the background contributions
may affect the associated production of the top quark pair and
Higgs boson has been discussed in
\cite{KS} for a few selected reactions (\ref{ee8f}).
A similar issue was also discussed in \cite{Moretti}, where
processes of the form $\epm \ra b\bar b b\bar b W^+W^- \ra b\bar b b\bar b
l^{\pm}\nu_lq\bar q'$ accounting for the signal of associated Higgs boson
and top quark pair production, as well as several irreducible background
reactions, were studied and in \cite{Schwinn}, where
pure electroweak (EW) contributions to
the leptonic and semileptonic reactions (\ref{ee8f}) were computed.
The off resonance background contributions in
$\epm \ra H b \bar b u\bar d \mu^- \bar \nu_{\mu}$ 
have been calculated in \cite{ustron07}.

In the present work, we will present results for the lowest order
cross sections of all reactions (\ref{ee8f}) possible in the SM calculated
with the complete set of the Feynman diagrams
with cuts on angles and energies of the final state particles
that should allow for identification of the corresponding jets and/or 
separate charged leptons \cite{expfeas}. Such results for reaction
(\ref{udmn}) were already shown in our former work \cite{KS}.
The size of the background contributions can be then inferred from
comparison of these cross sections with the signal cross sections
of the associated production of the top quark pair and Higgs boson.
To illustrate the size of the QCD background we present also the cross
sections calculated with the neglect of the gluon exchange contributions.
The calculation has been performed with {\tt carlomat}, a recently released
program for automatic computation of lowest order cross 
sections \cite{carlomat}. For testing purposes, the lowest order
cross sections calculated with the complete set of the Feynman diagrams
have been also computed with another multipurpose Monte Carlo (MC) generator 
{\tt WHIZARD/OMEGA} \cite{WHIZARD}. 
Moreover, we will include the initial state
radiation (ISR) effects and discuss the dependence of the cross sections
on the Higgs boson and top quark masses.

\section{Calculational details}

The results presented in Section~3 have been mostly obtained
with {\tt carlomat} \cite{carlomat}, a new multipurpose program for 
automatic computation of the lowest order cross sections whose main goal 
is to provide the reliable description of multiparticle reactions. 
Both {\tt carlomat} and the routines for the MC computation of cross
sections it generates are written in Fortran 90/95. 

The routine for matrix element calculation utilizes  the
helicity amplitude method.
Poles in the propagators of unstable particles
are regularized with constant particle widths
$\Gamma_a$  which are introduced through 
the complex mass parameters $M_a^2$ by making the substitution
\beq
\label{m2}
m_a^2 \;\ra \; M_a^2=m_a^2-im_a\Gamma_a, \qquad a=Z, W, H, t.
\eeq
Substitution (\ref{m2}) is made both in the $s$- and $t$-channel propagators.
The calculation is performed in
the complex mass scheme (CMS) \cite{Racoon}, where the electroweak mixing 
parameter $\sin^2\theta_W$ is defined as a complex quantity
\beq
\label{csw2}
\sin^2\theta_W=1-\frac{M_W^2}{M_Z^2},
\eeq
with ${M_W^2}$ and ${M_Z^2}$ given by (\ref{m2}).
The widths $\Gamma_a$ of (\ref{m2}), except for $\Gamma_Z$, whose actual 
value is rather
irrelevant in the context of associated top quark pair and Higgs boson
production and decay, are calculated in the lowest order of SM.
This, combined with the use of CMS, which preserves the Ward or 
Slavnov--Taylor identities and hence the gauge invariance, 
should minimize the unitarity violation effects. The latter are related
to the fact that substitution (\ref{m2}) introduces 
spurious terms of order $\mathcal{O}(\Gamma_a/m_a)=\mathcal{O}(\alpha)$ 
relative to the lowest order in places other than the resonant propagators. 
These higher order unitarity violating terms cannot be enhanced
due to the exact preservation of the Ward or Slavnov--Taylor identities, 
however, the use of other widths in (\ref{m2}) than the lowest order ones 
would have increased them, as we do not include any higher order corrections
to the unstable particle decays.

As in the next to leading order the number of the Feynman diagrams increases
dramatically, calculation of the full $\cal{O}(\alpha)$ radiative corrections
to any of reactions (\ref{ee8f}) is not feasible either at the moment or
even in the foreseeable future. However, precision of theoretical 
predictions for such reactions can be improved by the inclusion of
the ISR effects in the leading logarithmic (LL)
approximation in the structure function approach. 
The most recent version of {\tt carlomat} allows to compute
the ISR corrected
cross section in LL approximation ${\rm d} \sigma^{LL}(p_1,p_2)$ according
to the following formula
\bea
\label{LL}
  {\rm d} \sigma_{\rm Born + ISR}\left(p_1,p_2\right)=
\int_0^1 {\rm d} x_1 \int_0^1 {\rm d} x_2 \,
         \Gamma_{ee}^{LL}\left(x_1,Q^2\right)
\Gamma_{ee}^{LL}\left(x_2,Q^2\right) 
{\rm d}\sigma_{\rm Born}\left(x_1 p_1,x_2 p_2\right),
\eea
where $p_1$ ($p_2$) is the four momentum of a positron (electron),
$x_1$ ($x_2$) is the fraction of the initial momentum of the positron 
(electron) that remains after emission of a collinear photon and 
${\rm d} \sigma_{\rm Born}(x_1 p_1,x_2 p_2)$ is the lowest order cross 
section calculated at the reduced four momenta of the positron and electron. 
The structure function $\Gamma_{ee}^{LL}\left(x,Q^2\right)$ is given
by Eq.~(67) of \cite{Beenakker}, with {\tt `BETA'} choice for non-leading
terms. The splitting scale $Q^2$, which is not fixed in the LL approximation
is selected to be equal $s=(p_1+p_2)^2$.

{\tt carlomat} generates also dedicated phase space 
parametrizations which take into account mappings of peaks in the matrix 
element caused by propagators of massive unstable particles, of a photon, 
or a gluon in each Feynman diagram. This means that a number of different
phase space parametrizations generated is equal to a number of the Feynman 
diagrams. The phase space parametrizations 
are automatically implemented into a multichannel MC integration routine that 
performs integration over a 20-dimensional phase space of reactions
(\ref{ee8f}).
The integration is performed in several iterations, with the weights of 
different kinematical channels determined anew after each iteration. 
In this way, the kinematical channels with small weights are 
effectively not used in the subsequent iterations.

\section{Results}

We use the following set of initial physical parameters: 
the Fermi coupling, fine structure constant in the Thomson limit and strong
coupling
\bea
\label{params1}
G_{\mu}=1.16639 \times 10^{-5}\;{\rm GeV}^{-2}, \qquad
\alpha_0=1/137.0359991, \qquad \alpha_s(m_Z)=0.1176,
\eea
the $W$- and $Z$-boson masses
\bea
\label{vmass}
m_W=80.419\; {\rm GeV},\qquad m_Z=91.1882\; {\rm GeV},
\eea
the top quark mass and the heavy external fermion masses 
\beq
m_t=174.3\;{\rm GeV}, \quad m_b=4.8\;{\rm GeV}, \quad m_c=1.3\;{\rm GeV}, 
\quad m_{\tau}=1.77699\;{\rm GeV}. 
\eeq
The same masses are used both in the matrix elements and phase space
parametrizations. We would like to stress that, in the CMS, 
the complex top quark mass 
calculated according to Eq.~(\ref{m2}) enters both the top quark propagator,
including the numerator, as well as the top--Higgs Yukawa coupling.
In order to speed up the computation fermion masses smaller than those
of the $c$ quark and $\tau$ lepton have been neglected. 

The value of the Higgs boson mass is assumed at $m_H=130$~GeV.
To avoid unitarity violation the widths of $t$-quark, $W$- and Higgs bosons
are calculated to the lowest order of SM 
resulting in the following values:
\beq
\label{widths}
\Gamma_t=1.53088\;{\rm GeV},\qquad \Gamma_W=2.04764\;{\rm GeV},\qquad
\Gamma_H=8.0540\;{\rm MeV}.
\eeq
The $Z$ boson width, whose actual value is not relevant in the calculation, is
fixed at its experimental value $\Gamma_Z=2.4952$~GeV. 

We identify jets with their original partons and define the following 
basic cuts which should allow to detect events with separate 
jets and/or isolated charged leptons:
\begin{itemize}
\item cuts on an angle between a quark and a beam, an angle between two quarks
and on a quark energy: 
\beq
\label{cutsqq}
5^{\circ} < \theta (q,\mathrm{beam}) < 175^{\circ}, \qquad
\theta (q,q') > 10^{\circ}, \qquad E_{q} > 15\;{\rm GeV},
\eeq
\item cuts on angles between a charged lepton and a beam, a charged lepton 
and a quark and on energy of the charged lepton, $l=\mu,\tau$,
in the semileptonic detection channels of (\ref{ee8f})
\beq
\label{cutslq}
5^{\circ} < \theta (l,\mathrm{beam}) < 175^{\circ}, \qquad
\theta (l,q) > 10^{\circ}, \qquad E_l > 15\;{\rm GeV}, 
\eeq
\item a cut on an angle between the two charged leptons 
in the leptonic detection channels of (\ref{ee8f})
\beq
\label{cutsll}
\theta (l,l') > 10^{\circ},
\eeq
\item a cut on the missing transverse energy in the hadronic and semileptonic
detection channels of (\ref{ee8f})
\beq
\label{cutslqt}
/\!\!\!\!E^T > 15\;{\rm GeV}.
\eeq
\end{itemize}

The results for the lowest order cross sections of
the different channels of (\ref{ee8f}), which correspond to different decay
modes of the $W^+$ and $W^-$ bosons resulting from $t$ and $\bar t$ decays,  
are shown in Tables~\ref{Tab:lept}--\ref{Tab:had} for $\sqrt{s}=$
500~GeV, 800~GeV, 1~TeV and 2~TeV, the first two centre of mass energies 
being characteristic for the International Linear Collider (ILC) project 
\cite{ILC}  and the second two for the Compact Linear Collider (CLIC) design
\cite{CLIC}. We present cross sections of specific reactions, calculated 
with the complete set of the lowest order Feynmamn diagrams:
$\sigma^{\mathrm{Whiz.}}_{\mathrm{all}}$, 
computed with {\tt WHIZARD/OMEGA} and $\sigma_{\mathrm{all}}$, computed 
with {\tt carlomat}. Both $\sigma^{\mathrm{Whiz.}}_{\mathrm{all}}$
and $\sigma_{\mathrm{all}}$ have been computed with the same cuts and initial
input parameters. The comparison shows agreement within 
1--3 standard deviations of the MC integration that is indicated
in parentheses. In order to illustrate the size of the QCD 
background for each reaction we present also the cross sections 
$\sigma_{\mathrm{no\;QCD}}$ with the neglect of the gluon exchange 
contributions, computed with {\tt carlomat}.
Comparison of $\sigma_{\mathrm{all}}$ and
$\sigma_{\mathrm{no\;QCD}}$ shows that the QCD contributions are sizeable,
amazingly enough also for the leptonic channels of Table~\ref{Tab:lept}.
They are particularly large at $\sqrt{s}=500$~GeV, where the signal
is reduced because of the limited phase space volume available for 
reaction (\ref{eetth}). However, the QCD background is sizeable also
for higher centre of mass energies.
This is somewhat surprising if one takes 
into account a possibly low virtuality of the exchanged
gluons. Let us note that, despite quite different 
numbers of the Feynman diagrams contributing, the cross sections 
in Tables~\ref{Tab:lept}--\ref{Tab:had} are practically flavor
independent within each of the detection channels. This
means that, due to the cuts enforced, there is practically no dependence 
on $m_b$, $m_c$ and $m_{\tau}$ in the gluon and/or photon exchange 
background contributions.

The full size of the background contributions can be inferred from
comparison of the cross sections $\sigma_{\mathrm{all}}$ of
Tables~\ref{Tab:lept}--\ref{Tab:had} with the
corresponding signal cross sections presented in Table~\ref{Tab:sig}.
The background contributions are large, but they can be efficiently
reduced by imposing cuts on invariant masses allowing for
reconstruction of the top, antitop and Higgs boson. In particular, a narrow
cut, of the order of 1~GeV, on the invariant mass of the two $b$-jets that 
reconstruct the Higgs boson mass reduces the background very efficiently.
This issue has been
already extensively discussed in \cite{KS} so we will not address it here.
The signal cross sections are also the same for all the channels
within a given type: leptonic, semileptonic and hadronic one, as 
we have neglected the light fermion masses. However, even if the
light fermion masses had been kept non zero, the flavor dependence
of the cross sections would have been negligibly small, as amplitudes 
of the signal
diagrams of the associated production of the top quark pair and Higgs boson
are almost completely 
independent of the external fermion masses which always couple
to the propagators of the heavy particles.

Let us note that, in spite of somewhat different cuts imposed,
the cross sections of the semileptonic channels in 
Tables~\ref{Tab:lept}--\ref{Tab:sig} are roughly a factor
3 bigger than those of the leptonic channels. A similar relation
holds between the cross sections of the hadronic and semileptonic channels.

\begin{table}[!ht]
\begin{center}
\begin{tabular}{c|ccc|ccc}
\hline 
\hline 
\rule{0mm}{7mm}
&\multicolumn{3}{c|}{
$\epm\ra b \bar{b} b \bar{b} \nu_{e} e^+ \mu^- \bar{\nu}_{\mu} $}
&\multicolumn{3}{c}{
$\epm\ra b \bar{b} b \bar{b} \nu_{e} e^+ \tau^- \bar{\nu}_{\tau}$}
                                        \\ [1.5mm]
\hline 
\rule{0mm}{7mm}
No. of graphs: & -- & 37 868 & 33 500 & -- & 42 428 & 37 708\\ [1.5mm]
\hline 
\rule{0mm}{7mm}
$\sqrt{s}$ [GeV]& $\sigma^{\mathrm{Whiz.}}_{\mathrm{all}}$ 
& $\sigma_{\mathrm{all}}$ & $\sigma_{\mathrm{no\;QCD}}$
 & $\sigma^{\mathrm{Whiz.}}_{\mathrm{all}}$  
& $\sigma_{\mathrm{all}}$ & $\sigma_{\mathrm{no\;QCD}}$ \\[1.5mm] 
\hline
\rule{0mm}{7mm}
500  & 8.71(15)& 8.26(6) & 2.889(7) & 8.47(5) & 8.28(6) & 2.882(7)\\[1.5mm] 
800  & 36.2(1) & 35.6(2) & 24.41(4) & 36.1(1) & 36.0(1) & 24.42(4)\\[1.5mm] 
1000 & 34.1(1) & 34.3(2) & 22.50(4) & 34.2(1) & 34.3(1) & 22.57(5)\\[1.5mm] 
2000 & 18.6(2) & 18.4(2) & 11.15(7) & 18.2(2) & 18.3(1) & 11.06(7)\\[1.5mm] 
\hline 
\hline 
\rule{0mm}{7mm}
&\multicolumn{3}{c|}{
$\epm\ra b \bar{b} b \bar{b} \nu_{e} e^+ e^- \bar{\nu}_{e}$}
&\multicolumn{3}{c}{
$\epm\ra b \bar{b} b \bar{b} \nu_{\mu}\mu^+ \mu^- \bar{\nu}_{\mu}$}
                                        \\ [1.5mm]
\hline 
\rule{0mm}{7mm}
No. of graphs: & -- & 131 648 & 115 200 & -- & 46 890 & 40850 \\ [1.5mm]
\hline 
\rule{0mm}{7mm}
$\sqrt{s}$ [GeV]& $\sigma^{\mathrm{Whiz.}}_{\mathrm{all}}$ 
& $\sigma_{\mathrm{all}}$ & $\sigma_{\mathrm{no\;QCD}}$
 & $\sigma^{\mathrm{Whiz.}}_{\mathrm{all}}$  
& $\sigma_{\mathrm{all}}$ & $\sigma_{\mathrm{no\;QCD}}$ \\[1.5mm] 
\hline
\rule{0mm}{7mm}
500  & 8.43(4) & 8.43(6) & 2.895(8) & 8.43(1) & 8.49(7) & 2.890(8) \\[1.5mm] 
800  & 36.1(1) & 36.1(1) & 24.57(4) & 35.8(1) & 35.9(1) & 24.48(4) \\[1.5mm] 
1000 & 34.2(1) & 34.3(1) & 22.55(4) & 34.3(2) & 34.0(1) & 22.54(4) \\[1.5mm] 
2000 & 19.3(2) & 20.0(2) & 12.28(14)& 17.6(1) & 17.7(1) & 10.52(3) \\[1.5mm] 
\hline 
\hline 
\rule{0mm}{7mm}
&\multicolumn{3}{c|}{
$\epm\ra b \bar{b} b \bar{b} \nu_{\tau} \tau^+ \tau^- \bar{\nu}_{\tau}$}
&\multicolumn{3}{c}{
$\epm\ra b \bar{b} b \bar{b} \nu_{\mu}\mu^+ \tau^- \bar{\nu}_{\tau}$}
                                        \\ [1.5mm]
\hline 
\rule{0mm}{7mm}
No. of graphs: & -- & 66 658 & 58 666 & -- & 21 214 & 18 854 \\ [1.5mm]
\hline 
\rule{0mm}{7mm}
$\sqrt{s}$ [GeV]& $\sigma^{\mathrm{Whiz.}}_{\mathrm{all}}$ 
& $\sigma_{\mathrm{all}}$ & $\sigma_{\mathrm{no\;QCD}}$
 & $\sigma^{\mathrm{Whiz.}}_{\mathrm{all}}$  
& $\sigma_{\mathrm{all}}$ & $\sigma_{\mathrm{no\;QCD}}$ \\[1.5mm] 
\hline
\rule{0mm}{7mm}
500  & 8.44(1) & 8.43(5) & 2.907(6) & 8.40(3) &  8.46(3) & 2.885(6) \\[1.5mm] 
800  & 35.8(1) & 35.8(1) & 24.49(4) & 36.2(2) &  35.9(1) & 24.49(4) \\[1.5mm] 
1000 & 34.0(2) & 34.1(1) & 22.48(4) & 33.8(1) &  34.2(1) & 22.42(4) \\[1.5mm] 
2000 & 17.5(1) & 17.7(1) & 10.50(3) & 17.3(1) &  17.7(1) & 10.56(4) \\[1.5mm] 
\end{tabular} 
\end{center}
\caption{Cross sections in ab of leptonic detection channels of (\ref{ee8f}) 
calculated with {\tt WHIZARD/OMEGA} (first column) and {\tt carlomat} (second
column) with cuts given by (\ref{cutsqq}), (\ref{cutslq}) and (\ref{cutsll}).
The numbers in parenthesis show the MC uncertainty of the last decimal.}
\label{Tab:lept}
\end{table}
\begin{table}[!ht]
\begin{center}
\begin{tabular}{c|ccc|ccc}
\hline 
\hline 
\rule{0mm}{7mm}
&\multicolumn{3}{c|}{
$\epm\ra b \bar{b} b \bar{b} u \bar{d} e^- \bar{\nu}_{e}$}
&\multicolumn{3}{c}{
$\epm\ra b \bar{b} b \bar{b} c \bar{s} e^- \bar{\nu}_{e}$}
                                        \\ [1.5mm]
\hline 
\rule{0mm}{7mm}
No. of graphs: & -- & 53 632 & 39 224 & -- & 59 144 & 43 816 \\ [1.5mm]
\hline 
\rule{0mm}{7mm}
$\sqrt{s}$ [GeV]& $\sigma^{\mathrm{Whiz.}}_{\mathrm{all}}$ 
& $\sigma_{\mathrm{all}}$ & $\sigma_{\mathrm{no\;QCD}}$
 & $\sigma^{\mathrm{Whiz.}}_{\mathrm{all}}$  
& $\sigma_{\mathrm{all}}$ & $\sigma_{\mathrm{no\;QCD}}$ \\[1.5mm] 
\hline
\rule{0mm}{7mm}
500  & 26,8(1) & 26.9(1)  & 7.91(2) & 26.8(1) & 26.9(1) & 7.93(2) \\[1.5mm] 
800  & 99,9(3) & 100.6(3) & 67.3(1) &99.7(5)  & 100.5(3)& 67.1(1) \\[1.5mm] 
1000 & 94,0(3) & 95.0(3)  & 61.9(1) &92.4(4)  & 94.2(3) & 61.9(1) \\[1.5mm] 
2000 & 49.6(7) & 49.1(3)  & 29.4(1) &47.8(4)  & 48.2(3) & 29.8(2) \\[1.5mm] 
\hline 
\hline 
\rule{0mm}{7mm}
&\multicolumn{3}{c|}{
$\epm\ra b \bar{b} b \bar{b} u \bar{d} \mu^- \bar{\nu}_{\mu}$}
&\multicolumn{3}{c}{
$\epm\ra b \bar{b} b \bar{b} c \bar{s} \mu^- \bar{\nu}_{\mu}$}\\ [1.5mm]
\hline 
\rule{0mm}{7mm}
No. of graphs: & -- & 26 816 & 19 612 & -- & 29 572 & 21 908 \\ [1.5mm]
\hline 
\rule{0mm}{7mm}
$\sqrt{s}$ [GeV]& $\sigma^{\mathrm{Whiz.}}_{\mathrm{all}}$ 
& $\sigma_{\mathrm{all}}$ & $\sigma_{\mathrm{no\;QCD}}$
 & $\sigma^{\mathrm{Whiz.}}_{\mathrm{all}}$  
& $\sigma_{\mathrm{all}}$ & $\sigma_{\mathrm{no\;QCD}}$ \\[1.5mm] 
\hline
\rule{0mm}{7mm}
500  & 26.6(1)  & 26.9(1) & 7.91(2) & 26.5(1) & 26.6(1) & 7.89(2) \\[1.5mm] 
800  & 98.6(3)  & 100.6(2)& 67.0(1) & 98.4(3) & 100.0(3)& 67.3(1) \\[1.5mm] 
1000 & 93.3(3)  & 94.4(3) & 61.5(1) & 92.4(5) & 93.8(3) & 61.5(1) \\[1.5mm] 
2000 & 46.7(2)  & 47.6(2) & 28.3(1) & 47.0(1) & 46.9(2) & 28.2(1) \\[1.5mm] 
\hline 
\hline 
\rule{0mm}{7mm}
&\multicolumn{3}{c|}{
$\epm\ra b \bar{b} b \bar{b} u \bar{d} \tau^- \bar{\nu}_{\tau}$}
&\multicolumn{3}{c}{
$\epm\ra b \bar{b} b \bar{b} c \bar{s} \tau^- \bar{\nu}_{\tau}$}\\ [1.5mm]
\hline 
\rule{0mm}{7mm}
No. of graphs: & -- & 29 480 & 21 908 & -- & 32 332 & 24 300 \\ [1.5mm]
\hline 
\rule{0mm}{7mm}
$\sqrt{s}$ [GeV]& $\sigma^{\mathrm{Whiz.}}_{\mathrm{all}}$ 
& $\sigma_{\mathrm{all}}$ & $\sigma_{\mathrm{no\;QCD}}$
 & $\sigma^{\mathrm{Whiz.}}_{\mathrm{all}}$  
& $\sigma_{\mathrm{all}}$ & $\sigma_{\mathrm{no\;QCD}}$ \\[1.5mm] 
\hline
\rule{0mm}{7mm}
500  & 26.8(1) & 27.0(1) & 7.89(2) & 26.6(1)  &  26.6(1) & 7.90(2) \\[1.5mm] 
800  & 99.4(6) & 100.7(2)& 67.3(1) & 100.3(2) & 100.9(3) & 67.3(1) \\[1.5mm] 
1000 & 93.5(3) & 94.1(2) & 61.6(1) & 92.8(2)  & 94.0(3)  & 61.5(1) \\[1.5mm] 
2000 & 47.4(3) & 47.3(2) & 28.3(1) & 47.2(1)  & 47.0(2)  & 28.3(1) \\[1.5mm] 
\end{tabular} 
\end{center}
\caption{Cross sections in ab of semileptonic detection channels 
of (\ref{ee8f}) 
calculated with {\tt WHIZARD/OMEGA} (first column) and {\tt carlomat} (second
column) with cuts given by (\ref{cutsqq}), (\ref{cutslq}) and (\ref{cutslqt}).
The numbers in parenthesis show the MC uncertainty of the last decimal.}
\label{Tab:semilept}
\end{table}
\begin{table}[!ht]
\begin{center}
\begin{tabular}{c|ccc|ccc}
\hline 
\hline 
\rule{0mm}{7mm}
&\multicolumn{3}{c|}{
$\epm\ra b \bar{b} b \bar{b} u \bar{d} d \bar{u}$} &\multicolumn{3}{c}{
$\epm\ra b \bar{b} b \bar{b} c \bar{s} s \bar{c}$}\\ [1.5mm]
\hline 
\rule{0mm}{7mm}
No. of graphs: & -- & 185 074 & 81 250 & -- & 240 966 & 114 190 
\\[1.5mm]
\hline 
\rule{0mm}{7mm}
$\sqrt{s}$ [GeV]& $\sigma^{\mathrm{Whiz.}}_{\mathrm{all}}$ 
& $\sigma_{\mathrm{all}}$ & $\sigma_{\mathrm{no\;QCD}}$
& $\sigma^{\mathrm{Whiz.}}_{\mathrm{all}}$  
& $\sigma_{\mathrm{all}}$ & $\sigma_{\mathrm{no\;QCD}}$ \\[1.5mm] 
\hline
\rule{0mm}{7mm}
500  & 92.8(4)  & 94.1(3)  & 24.10(7)& 93.5(6) & 93.1(2) & 24.16(5) \\[1.5mm] 
800  & 318(2)   & 314(1)   & 205.7(3)& 308(2)  & 314(1)  & 205.7(3) \\[1.5mm] 
1000 & 284(2)   & 291(1)   & 187.4(3)& 286(2)  & 291(1)  & 187.6(3) \\[1.5mm] 
2000 & 137.1(5) & 139.2(7) & 83.1(3) & 137(1)  & 139.2(5)& 83.0(3)  \\[1.5mm]
\hline
\hline
\end{tabular}
\begin{tabular}{c|ccc}
\rule{0mm}{7mm}
&\multicolumn{3}{c}{
$\epm\ra b \bar{b} b \bar{b} u \bar{d} s \bar{c}$}\\ [1.5mm]
\hline 
\rule{0mm}{7mm}
No. of graphs: & -- & 39 342 & 25 246 \\ [1.5mm]
\hline 
\rule{0mm}{7mm}
$\sqrt{s}$ [GeV]&  $\sigma^{\mathrm{Whiz.}}_{\mathrm{all}}$  
& $\sigma_{\mathrm{all}}$ & $\sigma_{\mathrm{no\;QCD}}$ \\[1.5mm] 
\hline
\rule{0mm}{7mm}
500  & 92.2(3) &  93.2(2) & 24.10(5) \\[1.5mm] 
800  & 311(2)  &  314(2)  & 205.6(3) \\[1.5mm] 
1000 & 288(2)  &  290(1)  & 187.2(3) \\[1.5mm] 
2000 & 139(1)  &  139(1)  & 82.8(2)  \\[1.5mm] 
\end{tabular} 
\end{center}
\caption{Cross sections in ab of hadronic detection channels of (\ref{ee8f}) 
calculated with {\tt WHIZARD/OMEGA} (first column) and {\tt carlomat} (second
column) with cuts given by (\ref{cutsqq}) and (\ref{cutslqt}).
The numbers in parenthesis show the MC uncertainty of the last decimal.}
\label{Tab:had}
\end{table}
\begin{table}[!ht]
\begin{center}
\begin{tabular}{c|ccc}
\hline 
\hline 
\rule{0mm}{7mm}
$\sqrt{s}$ [GeV]&  $\sigma_{\mathrm{leptonic}}$  
& $\sigma_{\mathrm{semileptonic}}$ & $\sigma_{\mathrm{hadronic}}$ \\[1.5mm] 
\hline
\rule{0mm}{7mm}
500  & 1.12(1) & 3.095(3)  & 9.40(1) \\[1.5mm] 
800  & 16.86(4) & 46.27(2) & 141.4(1) \\[1.5mm] 
1000 & 14.67(4) & 40.18(2) & 122.5(1) \\[1.5mm] 
2000 & 5.67(7) & 15.14(3) & 44.58(3) \\[1.5mm] 
\end{tabular} 
\end{center}
\caption{Signal cross sections in ab of different 
detection channels of (\ref{ee8f}). The cuts are given by 
(\ref{cutsqq}), (\ref{cutslq}) and (\ref{cutsll}) for the leptonic, 
by (\ref{cutsqq}), (\ref{cutslq}) and (\ref{cutslqt}) for the 
semileptonic and by (\ref{cutsqq}) and (\ref{cutslqt}) for the hadronic 
channels. The numbers in parenthesis show the MC uncertainty of the last 
decimal.}
\label{Tab:sig}
\end{table}

It is interesting to look at the dependence of the cross sections on 
the Higgs boson and top quark masses. In order to illustrate this
we show in Table~\ref{Tab:masdep} the cross sections $\sigma_{\mathrm{all}}$,
$\sigma_{\mathrm{no\;QCD}}$ and $\sigma_{\mathrm{signal}}$ 
of reaction (\ref{udmn}) with cuts given by (\ref{cutsqq}), (\ref{cutslq}) 
and (\ref{cutslqt})
that have been calculated with 2 different hypothetical values of 
the Higgs boson mass: $m_H=115$~GeV and $m_H=130$~GeV, and 3 values of
the top quark mass: $m_t=171$~GeV,
$m_t=174.3$~GeV and $m_t=176$~GeV. The top quark mass has
been chosen within roughly $2\sigma$ range of the combined Tevatron 
result $m_t=173.1\pm 1.3$~GeV \cite{mtTev}. The change of particle
masses affects the corresponding lowest order SM widths of (\ref{widths}) 
which become
\bea
\Gamma_t(171\;{\rm GeV})=1.43152\;{\rm GeV}, \;&\;&\;
\Gamma_t(176\;{\rm GeV})=1.58351\;{\rm GeV}, \nn\\
\Gamma_H(115\;{\rm GeV})\!\!&=&\!\! 6.0223\;{\rm MeV}.
\eea
The cross sections for $m_H=115$~GeV are bigger than those for 
$m_H=130$~GeV. The relative difference is largest at $\sqrt{s}=500$~GeV, 
where it amounts to a factor 3--4 for the signal cross section, a factor 2
for pure EW contributions and a factor 1.3--1.5 for the full cross section.
The change of a top quark mass in Table~\ref{Tab:masdep} is 
smaller than that of the Higgs mass, however, it has proportionally the same
effect on the cross section at $\sqrt{s}=500$~GeV.
This can be easily traced back to the phase space volume limitations that are 
much stronger for the heavier Higgs boson and top quark.
At higher energies, where the relative differences in cross sections 
are smaller, 
the mass dependence becomes more involved as
the phase space effect interferes with the shift of the maximum of the 
cross section dependence on $\sqrt{s}$ that is caused by changes in $m_H$ 
and/or $m_t$.
For example,
the signal cross section at $\sqrt{s}=800$~GeV for $m_H=115$~GeV
is about 50\% bigger than that for $m_H=130$~GeV, while the cross section
for $m_t=171$~GeV is smaler smaler than that for $m_t=176$~GeV.
\begin{table}[!ht]
\begin{center}
\begin{tabular}{cc|ccc|ccc}
\hline 
\hline 
\rule{0mm}{7mm}
&&\multicolumn{3}{c|}{$m_H=115$~GeV} &\multicolumn{3}{c}{
$m_H=130$~GeV}\\ [1.5mm]
\hline 
\rule{0mm}{7mm}
$\sqrt{s}$ [GeV] & $\;m_t$ [GeV] & 
$\sigma_{\mathrm{all}}$ & $\sigma_{\mathrm{no\;QCD}}$
& $\sigma_{\mathrm{signal}}$ & $\sigma_{\mathrm{all}}$ 
& $\sigma_{\mathrm{no\;QCD}}$ & $\sigma_{\mathrm{signal}}$ \\[1.5mm] 
\hline
\rule{0mm}{7mm}
500  & 171  & 39.8(1) & 20.21(5) & 14.61(1) & 29.9(1) & 10.29(3) & 4.670(4) \\
     & 174.3& 35.3(1) & 16.41(4) & 11.69(1) & 26.9(1) & 7.91(2) & 3.095(3) \\
     & 176  & 33.3(1) & 14.52(4) & 10.19(1) & 25.4(1) & 6.75(2) & 2.365(2) 
\\[1.5mm] 
800  & 171  &118.2(2) & 84.8(1) & 63.17(3) & 99.2(2) & 66.0(1) & 44.62(2) \\
     & 174.3&119.9(3) & 86.8(1) & 65.61(3) & 100.6(2)& 67.0(1) & 46.27(2) \\
     & 176  &120.3(3) & 88.0(1) & 66.83(3) & 100.4(2)& 67.8(1) & 47.08(2) 
\\[1.5mm] 
1000 &  171 &105.8(3) & 73.8(1) & 51.73(2) & 92.1(2) & 59.8(1) & 38.34(2) \\
     & 174.3&108.6(3) & 75.7(1) & 54.22(3) & 94.4(3) & 61.5(1) & 40.18(2) \\
     & 176  &110.2(3) & 77.7(1) & 55.51(3) & 94.8(2) & 62.3(1) & 41.07(2) 
\\[1.5mm] 
2000 &  171 & 49.9(2) & 31.37(9) & 18.17(5) & 46.2(2) & 27.31(7) & 14.34(4) \\
     & 174.3& 51.5(2) & 32.64(9) & 19.20(5) & 47.6(2) & 28.35(8) & 15.14(3) \\
     & 176  & 51.8(2) & 33.21(9) & 19.78(5) & 47.5(2) & 28.75(9) & 15.60(3) 
\\[1.5mm] 
\end{tabular} 
\end{center}
\caption{Cross sections in ab of (\ref{udmn})
calculated with {\tt carlomat} with different values of the top quark mass
$m_t$ and the Higgs boson mass $m_H$
with cuts given by (\ref{cutsqq}), (\ref{cutslq}) and (\ref{cutslqt}).
The numbers in parenthesis show the MC uncertainty of the last decimal.}
\label{Tab:masdep}
\end{table}

In order to illustrate the potential effect that the unitarity 
violating terms discussed in Section~2 may have on the presented results 
we show 
in Table ~\ref{Tab:CMSvFWS} the lowest order cross sections of (\ref{udmn}) 
calculated with {\tt carlomat} for two different values of the Higgs boson 
mass $m_H$ and $m_t=174.3$ GeV 
in the CMS and fixed width scheme (FWS). The latter differs from the CMS
in the definition of the EW mixing parameter that instead of 
by Eq. (\ref{csw2}) is defined by
\beq
\label{sw2}
\sin^2\theta_W=1-\frac{m_W^2}{m_Z^2},
\eeq
with $m_W$ and $m_Z$ being physical masses of the EW bosons. 
As the cross
sections agree within one standard deviation of the MC integration, we see
that the unitarity violation terms are not enhanced 
in the FWS at the presented values of $\sqrt{s}$, even though the Ward 
or Slavnov--Taylor identites are not stisfied in this scheme.

\begin{table}[!ht]
\begin{center}
\begin{tabular}{c|cc|cc}
\hline 
\hline 
\rule{0mm}{7mm}
&\multicolumn{2}{c|}{$m_H=115$~GeV} 
&\multicolumn{2}{c}{$m_H=130$~GeV}\\ [1.5mm]
\hline 
\rule{0mm}{7mm}
$\sqrt{s}$ [GeV] 
& $\sigma_{\mathrm{all}}$ & $\sigma_{\mathrm{all}}^{\mathrm{FWS}}$
& $\sigma_{\mathrm{all}}$ & $\sigma_{\mathrm{all}}^{\mathrm{FWS}}$\\[1.5mm] 
\hline
\rule{0mm}{7mm}
500  & 35.3(1) & 35.4(1) & 26.9(1) & 26.8(1) \\[1.5mm] 
800  &119.9(3) & 120.3(3)& 100.6(2)& 100.8(3)\\[1.5mm] 
1000 &108.6(3) & 108.7(2)& 94.4(3) & 94.5(2) \\[1.5mm] 
2000 & 51.5(2) & 51.3(2) & 47.6(2) & 47.3(2) \\[1.5mm] 
\end{tabular} 
\end{center}
\caption{Cross sections in ab of (\ref{udmn}) in the CMS and FWS
calculated with {\tt carlomat} for two different values of the Higgs boson 
mass $m_H$ and $m_t=174.3$ GeV.
Cuts are given by (\ref{cutsqq}), (\ref{cutslq}) and (\ref{cutslqt}).
The numbers in parenthesis show the MC uncertainty of the last decimal.}
\label{Tab:CMSvFWS}
\end{table}

The impact of the ISR is illustrated in 
Table~\ref{Tab:ISR}, where
we show the cross sections $\sigma_{\mathrm{all+ISR}}$, 
$\sigma_{\mathrm{no\;QCD+ISR}}$ and $\sigma_{\mathrm{signal+ISR}}$ 
for reaction (\ref{udmn}) including 
ISR according to Eq.~(\ref{LL}) with the splitting scale $Q^2=s$. They 
have been calculated with 
the Higgs boson masses $m_H=115$~GeV and $m_H=130$~GeV, 
and the top quark mass $m_t=174.3$~GeV. 
These cross sections should be compared
with the Born cross sections at the corresponding centre of mass energy 
and $m_t=174.3$~GeV in Table~\ref{Tab:masdep}. We see that,
as could be expected from Eq.~(\ref{LL}), ISR decreases substantially
the signal and pure EW cross sections at $\sqrt{s}=500$~GeV, as due to 
the photon emission
the actual centre of mass energy of $\epm$ scattering may fall down below 
the $t\bar{t}H$ threshold, where the signal cross section 
$\sigma_{\mathrm{signal}}$ becomes very small. Note that the effect is smaller 
for $m_H=115$~GeV due to more phase space volume available. The decrease of 
the full cross section at this energy is relatively smaller, as it is 
dominated by the QCD background contributions. At 
$\sqrt{s}=800$~GeV and $\sqrt{s}=1$~TeV, 
the cross section reduction is much smaller 
and, at $\sqrt{s}=2$~TeV, there is 
even some growth in the ISR corrected cross sections with respect to the 
lowest order cross sections of Table~\ref{Tab:masdep}.

\begin{table}[!ht]
\begin{center}
\begin{tabular}{c|ccc|ccc}
\hline 
\hline 
\rule{0mm}{7mm}
&\multicolumn{3}{c|}{$m_H=115$~GeV} &\multicolumn{3}{c}{$m_H=130$~GeV}\\ [1.5mm]
\hline 
\rule{0mm}{7mm}
$\sqrt{s}$ [GeV] 
& $\sigma_{\mathrm{all + ISR}}$ & $\sigma_{\mathrm{no\;QCD + ISR}}$ 
                                         & $\sigma_{\mathrm{signal + ISR}}$ 
& $\sigma_{\mathrm{all + ISR}}$ & $\sigma_{\mathrm{no\;QCD + ISR}}$ 
                                         & $\sigma_{\mathrm{signal + ISR}}$ 
\\[1.5mm] 
\hline
\rule{0mm}{7mm}
500  & 29.2(1) & 12.24(4) & 8.48(2)  & 22.6(1) & 5.89(2) & 2.114(6) \\[1.5mm]
800  &114.1(2) & 82.2(1)  & 62.34(6) & 94.3(2) & 62.9(1) & 43.39(4) \\[1.5mm]
1000 &107.5(3) & 75.3(1)  & 54.29(5) & 91.9(2) & 60.5(1) & 39.73(4) \\[1.5mm]
2000 &54.7(2)  & 35.2(1)  & 21.17(5) & 50.0(2) & 30.2(1) & 16.50(3) \\[1.5mm]
\end{tabular} 
\end{center}
\caption{Cross sections in ab of (\ref{udmn}) including ISR
in the LL approximation
calculated with {\tt carlomat} for two different values of the Higgs boson 
mass $m_H$ and $m_t=174.3$ GeV.
Cuts are given by (\ref{cutsqq}), (\ref{cutslq}) and (\ref{cutslqt}).
The numbers in parenthesis show the MC uncertainty of the last decimal.}
\label{Tab:ISR}
\end{table}

\section{Summary}

We have presented results for the lowest order cross sections, calculated
with the complete set of the standard model Feynman diagrams, of all 
reactions (\ref{ee8f}) relevant for the associated production 
of the top quark pair and Higgs boson that can be used for determination 
of the top--Higgs Yukawa coupling at the $\epm$ linear collider. 
A comparison with the corresponding signal cross sections of (\ref{eetth})
has shown that the background contributions are large for typical
particle identification cuts.
In particular, the QCD background contributions are much bigger than it
could have been expected taking into account a possibly low virtuality 
of exchanged gluons. As we have shown elsewhere \cite{KS} for a few 
representative
channels of (\ref{ee8f}) the background can
be efficiently reduced by imposing invariant mass cuts allowing for
the top and antitop quark, and Higgs boson identification. 
We have shown that the unitarity 
violating terms discussed in Section~2 have practically no effect
on the lowest order cross sections of (\ref{udmn}) at the considered range of
$\sqrt{s}$ by comparing the cross sections calculated in the CMS and FWS,
where the latter violates gauge invariance.

Moreover, we have included the ISR and illustrated its effects
concentrating on a semileptonic reaction (\ref{udmn}) for which
we have have also discussed the dependence of the cross sections on 
the Higgs boson and top quark masses.
Taking into account cross sections of the different $t\bar{t}H$ detection
channels presented in 
Tables~\ref{Tab:lept}--\ref{Tab:ISR} the best place to measure the 
top--Higgs Yukawa coupling seem to be a linear collider operating at
the centre of mass energy of about 800~GeV.

Acknowledgements: This work is supported in part by the Polish Ministry 
of Science under Grant No. N N519 404034 and by European Community's 
Marie-Curie Research Training Network under contracts MRTN-CT-2006-035482 
(FLAVIAnet) and MRTN-CT-2006-035505 (HEPTOOLS).

\end{document}